\documentclass[prl,twocolumn,amsmath,amssymb,floatfix,showpacs,superscriptaddress]{revtex4}
\usepackage{graphicx}
\usepackage{amsmath}
\usepackage{bm}

\newcommand{\bq}{\begin{equation}}
\newcommand{\ee}{\end{equation}}
\newcommand{\fr}[2]{\frac{#1}{#2}}

\renewcommand{\vr}{\vec r}
\newcommand{\vR}{\vec R}

\renewcommand{\vec}[1]{\mathbf{#1}}
\usepackage{colordvi}
\usepackage{graphicx}
\usepackage{color}

\begin{document}
\title{Wigner crystal of a two-dimensional electron gas \\
with a strong spin-orbit interaction}
%Wigner Crystal in a Two-Dimensional Electron Gas with Strong
%Spin-Orbit Interaction}
\date{\today }

\author{P. G. Silvestrov}
\affiliation{Physics Department and Dahlem Center for Complex
Quantum Systems, Freie Universit\"{a}t Berlin,
%Arnimallee 14,
14195 Berlin, Germany} \affiliation{Institute for Mathematical
Physics, TU Braunschweig, 38106 Braunschweig, Germany}

\author{O. Entin-Wohlman}

\affiliation{Physics Department, Ben Gurion University,  Beer
Sheva 84105, Israel} \affiliation{Raymond and Beverly Sackler
School of Physics and Astronomy, Tel Aviv University, Tel Aviv
69978, Israel}

\begin{abstract}

The Wigner-crystal phase of two-dimensional electrons interacting
via the Coulomb repulsion and subject to a strong Rashba
spin-orbit coupling is investigated. For low enough electronic
densities the spin-orbit  band splitting can be larger than the
zero-point energy of the lattice vibrations. Then the degeneracy
of the lower subband results in a spontaneous symmetry breaking of
the  vibrational ground state. The $60^{\circ}-$rotational
symmetry of the triangular (spin-orbit coupling free) structure is
lost, and the unit cell of the new lattice contains two electrons.
Breaking the rotational symmetry also  leads  to  a (slight)
squeezing of the underlying triangular lattice.

\end{abstract}
 %73.20.Qt, Electron solids
 %75.70.Tj, Spin-orbit effects
\maketitle

%\noindent
{\it 1. Introduction.} The Wigner crystal~\cite{Wigner34}, the
insulating companion of a two-dimensional metal, is predicted to
appear in an electron gas of ultra-low densities formed  in
semiconductor heterostructures when the Coulomb repulsion-induced
ordering wins over the zero-point quantum fluctuations
\cite{Tanatar89,Drumond09}. Low densities amount to very clean
samples. That is why experimentally Wigner crystals were observed
either in naturally clean systems, like electrons on the surface
of liquid Helium~\cite{Helium79}, or in two-dimensional
semiconductors when the kinetic energy  is suppressed by a strong
magnetic field~\cite{EAndrey88},  or due to a large mass of the
charge carriers~\cite{Yoon99}. However, lowering the electronic
density not only enhances the electronic correlations, but also
tends to increase the relative importance of the spin-orbit
interaction, generically present in low-dimensional systems
\cite{Winkler}. Thus attempting to increase the role of electronic
repulsion may lead one into a regime in which quantum fluctuations
around the classical equilibrium sites of the Wigner crystal are
dominated by the spin-orbit interaction. Such crystals, as we
show, demonstrate a number of unexpected properties, having no
analogue hitherto.

\begin{figure}
\includegraphics[width=8.cm]{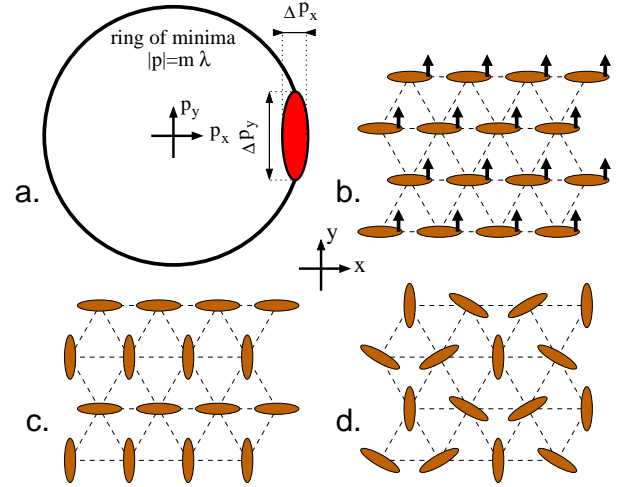}
\caption{\label{fig:1} Schematic visualization of the  electronic
density in a two-dimensional electron crystal. $a$~-~the ring of
minima in  momentum space for the lower Rashba subband
$E_{p-}=(p-m\lambda)^2/(2m)$. The colored ellipse shows the electronic
density (for  structure $b$). The density is centred around a
particular minimum, but is strongly elongated along the line of
minima, $\Delta p_x\ll\Delta p_y\ll m\lambda$, which helps to
lower the interaction energy. $b,c,d$~-~three possible periodic
configurations (see text). The dark ellipses indicate the
directions of electrons' oscillations. Structure $c$ has the
lowest zero-point energy of the lattice vibrations. The arrows in
$b$ indicate the in-plane spin orientations. } \label{fig1}
\end{figure}

The structure of a crystal is usually determined by the
interaction between particles which oscillate slightly around
their static equilibrium positions. For electrons subject to the
Rashba spin-orbit interaction~\cite{Rashba60}, which we consider
in this paper, this picture is modified, since in this case even
the notion of the ``static particle position" requires a
clarification. The spin-orbit interaction splits the electron
spectrum, leading to a Mexican hat shaped lower subband with a
circle of degenerate minima. However, in an interacting
quantum-mechanical system, different minima of the single-particle
spectrum are not equivalent.
%{\color{red} Different minima of the
%single-particle spectrum, however, are not equivalent in
%interacting quantum mechanical system.}
Picking up one of these minima breaks the rotational symmetry in
momentum space, and then, via the uncertainty relation, the
spatial motion and the Coulomb interaction of the electrons are
affected. It has been even recently shown~\cite{Berg12} that the
Rashba spin-orbit interaction stabilizes a strongly asymmetric
Wigner crystal when the interaction potential is short-range,
$V\sim 1/r^\alpha$ with $\alpha>2$. Such a crystal would have been
otherwise unstable for any small electronic density. In this paper
we analyze  the crystal created by electrons interacting via the
{\em unscreened} Coulomb repulsion, $V=e^{2}/r$. Then the
triangular lattice of the crystal~\cite{Bonsall79} remains stable
on the classical level. However, breaking the symmetry in %the
momentum plane changes drastically the fluctuation properties and
the electronic density distribution.

We consider strong spin-orbit interactions, such that the
effective  Hilbert space  is reduced to include   only electronic
states of momenta close to the ring of minima in the lower subband
(see Fig. \ref{fig1}$a$). Each electron in the crystal picks up
only one of those minima. However, since the electron's
displacement in  momentum space along the line of minima costs no
kinetic energy, one may effectively freeze its spatial vibrations
in this direction. Freezing the spatial vibrational mode reduces
the average potential energy, since now the electron never leaves
its classical equilibrium location in this direction. The
vibrations along the direction perpendicular to the ring of minima
have the same effective mass as in the absence of the spin-orbit
interaction. Hence the electrons' fluctuations when the spin-orbit
coupling is strong enough  are equivalent to vibrations of
particles having anisotropic masses. Here though,  the light and
the (infinitely) heavy masses directions are chosen for each
electron individually.

\begin{figure}
\includegraphics[width=8.cm]{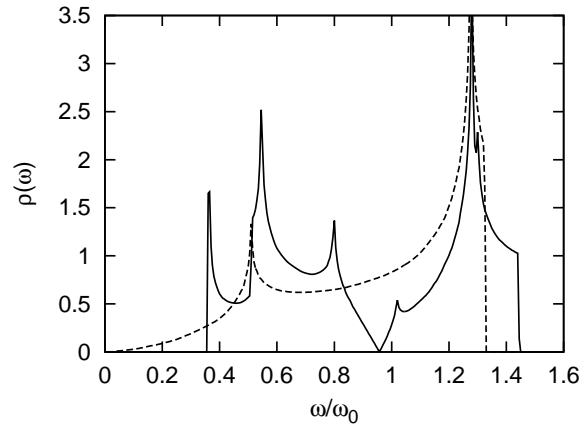}
\vspace{-.5cm} \caption{\label{fig:2} The vibration density of
states as a function of the frequency (scaled by $\omega_{0}$) for
the Wigner-crystal configuration of  Fig.  \ref{fig:1}$b$ (dashed
line) and of Fig. \ref{fig:1}$c$ (solid line).  Though the
spectrum extends all the way down to $\omega =0$  for $1b$ the
average frequency turns out to be smaller for $1c$. Note the
linear vanishing of the density (characteristic of a Dirac cone)
in the middle of spectrum for $1c$ and especially the inverse
square-root divergence of the density at the lower edge,
$\omega\approx 0.36\omega_0$.  }
%\vspace{-.3cm}
%\label{fig1}
\end{figure}

As different minima of the Rashba Hamiltonian are classically
equivalent, the proper choice of the electrons' configuration
should be the one minimizing  the zero-point fluctuation energy of
the crystal. Finding the minimum of the zero-point energy for
general directions of the oscillations is a difficult task which
will not be fully accomplished in this paper. Instead, we adopt a
step-by-step approach, considering  a series of configurations
depending on 1, 2, 3, $\ldots$, angles with respect to which one
would minimize the energy functional. The first three steps in
this scheme are illustrated in Figs.~\ref{fig:1}$b$, $c$, and $d$.
The dark ellipses there indicate the directions along which the
electrons vibrate. First, one requires all electrons to vibrate in
unison, and minimizes the energy with respect to a single angle.
Figure~\ref{fig:1}$b$ shows the best configuration of such a
one-parameter family. At the next step one allows  two neighboring
electrons on the triangular lattice to vibrate along independent
directions and then repeats this configuration periodically. The
unit cell now contains two electrons. Figure~\ref{fig:1}$c$ shows
the configuration realizing the minimum of the vibrational energy
for such a two-parameter family of crystals. One may consider
similarly  a lattice with more independent vibrational directions,
see Fig.~\ref{fig:1}$d$. Among all  crystal configurations which
we have analyzed, the one in Fig.~\ref{fig:1}$c$ has the smallest
vibrational energy. Different crystal configurations have also a
very different vibrational spectra [Fig.~\ref{fig:2} below] which
can be exploited to distinguish them experimentally.

%\noindent
{\it 2. Wigner crystal with spin-orbit coupling.} The Hamiltonian
of the system under consideration is
 \begin{align}
{\cal H}=\sum_i {\cal H}^{}_{0i}
+\sum_{i<j}e_{}^2/|\vR^{}_{ij}+\vr_{ij}^{}|\ , \label{Vij}
 \end{align}
with ${\bf R}_{ij}\equiv {\bf R}_{i}-{\bf R}_{j}$. At equilibrium,
the electrons are located at sites  $\vR_i$ on a triangular
lattice \cite{Bonsall79} of  spacing $a$. The oscillations around
those sites are described by  expanding the interaction in the
small displacements $|\vr_i|\ll a$ up to second order \cite{tech}. This
expansion yields a single-electron harmonic potential and a
(bi)linear in the displacement electron-electron interaction [see Eq.
(\ref{Hphonon}) below]. The former allows to introduce the
frequency~\cite{Flambaum99}
 \begin{align}\label{Omega0}
\omega^{}_{0}=\sqrt{\gamma e^{2}/(ma^{3})}\ ,
 \end{align}
with $\gamma=\sum_{i\neq 0}a^{3}/(2R^{3}_{i})=5.5171$ and $m$
being the effective mass. The  %effects we study stem from the
single-electron part of the Hamiltonian (\ref{Vij}) reads
 \begin{align}
 \label{Rashba}
{\cal H}^{}_0 ={\mathbf p^2}/(2m) +\lambda (\sigma_x p_y -\sigma_y p_x)
+m\lambda^2/2\ ,
 \end{align}
where $\lambda$ denotes the Rashba spin-orbit coupling strength
and $\sigma_{x,y} $ are the Pauli matrices. The spectrum of ${\cal
H}_0$ consists of  two subbands, $E_{p\pm} =(p\pm
m\lambda)^2/(2m)$, which correspond to electrons with in-plane
spins directed normal to the momentum, and pointing to its   left
or right, respectively. We focus on the regime where the
spin-orbit energy exceeds  the one due to the zero-point motion of
the electrons around their equilibrium sites,
 \begin{align}
 \label{regime}
m\lambda_{}^2 \gg \hbar\omega_0 \sim \sqrt{
e^2_{}\hbar^2_{}/(ma_{}^3)}\ ,
 \end{align}
which means that the electrons are always confined to the lowest
subband. The relative strength of the Coulomb interaction compared
to the kinetic energy is characterized by the dimensionless
parameter $r_s$, related to the electronic density $n$ and the
Bohr radius $a_{\rm B}=\hbar^2/(me^2)$ as $\pi r_s^2=1/(n a_{\rm
B}^2)$~\cite{rs}. Obviously, in the case of a strong Rashba
spin-orbit interaction the large value of the same parameter
ensures the existence of the Wigner crystal: $r_s\gg 1$ here means
that the gain in the Coulomb energy per electron in the ordered
phase exceeds the energy rise due to the zero-point fluctuations
of electrons in the lattice, $e^2/a\gg \hbar\omega_0$. (Quantum
Monte-Carlo simulations indicate that in the absence of spin-orbit
coupling the Wigner crystal exists at
$r_s>35$~\cite{Tanatar89,Drumond09}.)

Once the electrons reside in the lower Rashba subband, their
(ground-state) momenta   lie   within a narrow ring in  momentum
plane of radius $m\lambda$ and   width $\Delta p$,  determined by
the zero-point energy
 \begin{align}
 \label{ring}
(p-m\lambda)^2_{}\lesssim \Delta p^2_{} \sim\sqrt{
e^2_{}\hbar^2_{} m/a^3_{}} \ll (m\lambda)^2 _{}\ .
 \end{align}
However, in the crystal different parts of this ring are not
equivalent and each electron may choose its own small sector.
Imagine a much-elongated wave packet built from the lower Rashba
subband solutions, such that
 \begin{align}
 \label{WavePack}
| p^{}_x-m\lambda|\lesssim \Delta p^{}_x, \ | p^{}_y | \lesssim
\Delta p^{}_y , \ \Delta p_x\ll \Delta p_y\ll m\lambda ,
 \end{align}
as in Fig. \ref{fig1}$a$. The two spatial dimensions of this wave
packet are  very different, $\Delta x \gg \Delta y $, and hence
the minute displacement along the $y-$direction gives a negligible
correction to the interaction energy. The expectation value of
${\cal H}_{0}$, which has an eigenvalue $E_-=0$ at $|p|=m\lambda$,
is determined by the smaller momentum extension of the packet,
$\langle {\cal H}^{}_0\rangle \sim \Delta p_x^2/(2m)$. This
implies that   the system may choose an anisotropic  pattern where
around each lattice site the density forms a narrow ellipse of
length $\Delta x\sim [a^3_{}\hbar_{}^2/(m e^2)]^{1/4}$. The
orientations of these ellipses   will be determined by the
zero-point energy of the vibrations pertaining to a specific
pattern.

%\noindent
{\it 3. Vibration spectrum.} We begin with  the configuration
shown  in Fig.~\ref{fig:1}$b$. In this structure the electrons
oscillate along $x$, and (since the Hilbert space is reduced to
that of the lower Rashba subband) are strongly spin-polarized
along $ y$, $\langle\sigma_{y}\rangle \simeq 1$.
%, as shown  in Fig.~\ref{fig:1}$b$.
To find  the excitation spectrum around this
particular structure, we shift the $x-$component of the momentum,
$p_{x}\rightarrow m\lambda +p_{x}$, multiplying the many-electron
wave function by $\exp[(i/\hbar )\sum_{i} m\lambda x_i]$. The
reduced Hamiltonian contains only the term quadratic in $p_{x}$,
while $p_{y}$  appears  at higher orders in $p_{y}/(m\lambda)$ and
may be discarded. The uncertainty principle then imposes no
restrictions on the displacements along  the $y-$direction,
allowing to choose $y_i\equiv 0$. This results in a
single-coordinate effective Hamiltonian,
 \begin{align}
\label{Hphonon} {\cal H}^{}_{\mathrm {eff}} &=\sum_i
[p_{x_i}^2/(2m) +m\omega_0^2 x_i^2/2] +e_{}^2\sum_{i<j}
x^{}_ix^{}_j u_{ij}\ ,
 \end{align}
where $u^{}_{ij}=1/R_{ij}^3 -3X_{ij}^2/R_{ij}^5$, $X_{ij}$ is the
$x-$component of ${\bf R}_{ij}$ and $\omega_0$ is defined in
Eq.~(\ref{Omega0}). Hamiltonian Eq.~(\ref{Hphonon}) contains half
of the degrees of freedom of the original one, since it allows for
a single vibrational direction per electron. In the regime given
by Eq.~(\ref{regime}), the missing degrees of freedom pertain to %the
low energy non-phonon excitations, whose analysis is beyond the
scope of this paper.

Exploiting the Bogoliubov transformation
 \begin{align}
c^{}_{\bf k} \equiv \sum_{i}e^{-i{\bf k}\cdot{\bf R}_{i}}
\fr{m\omega^{}_0 x^{}_i+ip_{x_{i}}}{\sqrt{2Nm\omega^{}_0\hbar}} =
{\rm cosh} u^{}_{\bf k}d^{}_{\bf k}+{\rm sinh} u^{}_{\bf
k}d^{\dagger}_{-{\bf k}} ,
 \end{align}
where $N$ is the number of lattice sites, transforms the
Hamiltonian into
 \begin{align}
 \label{HphononBogolyub}
{\cal H}^{}_{\rm eff}=  \sum_{\bf k} \hbar\omega^{}_{\bf k}
(d_{\bf k}^\dagger d_{\bf k}^{} +\frac{1}{2})\ , \ \omega^{}_{\bf
k}=\omega_0 \sqrt{1+2v({\bf k})}\ .
 \end{align}
Here, $\tanh 2u_{\bf k}=-v({\bf k})/[1+v({\bf k})]$ and $v({\bf
k})= \sum_{i\neq 0}  e^{i{\bf k}\cdot {\mathbf R_{0i}}}
[a^3/(2\gamma R_{0i}^3)]\left(1- 3X_{0i}^2/R_{0i}^2\right)$. In
particular, at small wavevectors the dispersion law becomes
 \begin{align}
\label{Omega} \omega_{\bf k }^{}=[{4\pi
e^2}/({\sqrt{3}ma^2})]^{1/2}_{}{|k_x|}/{\sqrt{k}}\ .
 \end{align}
The $\sqrt{k}-$dependence of the low %long-wave
frequencies is expected for a two-dimensional plasmon gas. The
striking  feature though is the {\em angular dependence} of the
dispersion law (\ref{Omega}), with vanishing frequency along the
$y-$direction. It signals a spontaneous symmetry-breaking caused
by the degeneracy at the bottom of the lower  Rashba subband.
%{\color{red} degeneracy lifting?}

The vibration spectra for the structures with several electrons
per unit cell, like in Figs.~\ref{fig:1}$c$ and $d$ are found
similarly. Details of this calculations are given in
Ref.~\onlinecite{supplementary}.

%\noindent
{\it 4. Minimizing the zero-point energy.} We have investigated
numerically all possible configurations of the Wigner crystal with
vibrational directions depending on 1, 2, and 3 angles, as
described in the introduction.  Important examples representing
these families of configurations are shown in
Figs.~\ref{fig:1}$b$, $1c$, and $1d$. The ground-state energy is
determined by the frequency $\omega_{\vec k}$ averaged  over the
Brillouin zone, which in the case of Fig.~\ref{fig:1} gives
 \begin{align}\label{threeOmega}
\fr{\langle \omega_{\vec k}^b \rangle}{\omega_0} =0.951\  , \
\fr{\langle \omega_{\vec k}^c \rangle}{\omega_0} =0.939\  , \
\fr{\langle \omega_{\vec k}^d \rangle}{\omega_0} =0.971\  .
 \end{align}
Quite unexpectedly, among the examined configurations the one with
two electrons per unit cell shown in Fig.~\ref{fig:1}$c$ has the
smallest zero-point energy. Configuration 1$b$ has the smallest
$\langle \omega_{\vec k} \rangle$ for the families with 1 and 3
electrons per unit cell, but is not the global minimum of the
vibrational energy. The highly-symmetric configuration 1$d$ has
the largest zero-point energy in the family with 3 independent
vibrational directions.

All  phonon frequencies coincide independent of the direction of
vibration, $ \omega_{\bf k} \equiv\omega_{0}$, upon exploiting the
Einstein approximation, in which each electron is confined to a
harmonic potential created when the locations of all other
electrons are frozen \cite{tech}. In all considered crystal
configurations the average frequency was always close to the
Einstein approximation, like in Eq.~(\ref{threeOmega}).

One way to probe experimentally the structure of a crystal is to
measure the vibration spectrum. Figure~\ref{fig:2} shows the
density of states (DOS) as a function of the frequency for the
structures in Figs.~\ref{fig:1}$b$ and $1c$. Several interesting
features can be observed there (but unfortunately, none  explains
which structure is energetically favorable, and why the resulting
$\langle\omega_{\bf k}\rangle $'s are so close to each other). For
the configuration \ref{fig:1}$b$ the DOS is finite all the way
down to zero frequency due to the plasmon mode Eq.~(\ref{Omega}).
The spectrum has a step singularity at the high-frequency end, and
two logarithmic van Hove singularities. Though  the modes
describing global translations for the full Hamiltonian
Eq.~(\ref{Vij}) always have zero energy,  it is only for the
structure \ref{fig:1}$b$ that one of these  modes is reproduced by
the effective  Hamiltonian Eq.~(\ref{Hphonon}).

The vibration spectrum of the crystal phase shown  in
Fig.~\ref{fig:1}$c$ is gapped (the solid line in Fig.
\ref{fig:2}). The DOS has several step singularities and
logarithmic divergences. The linear vanishing of the DOS at
$\omega\approx \omega_0$ corresponds to a conical crossing of the
two bands (a Dirac point). Surprisingly enough, we observe an
almost perfectly inverted square-root singularity at low
frequencies, $\rho(\omega)\sim [\omega -0.36\omega_0]^{-1/2}$;
such a divergence usually characterizes {\em one-dimensional}
systems. Careful numerical investigation shows that  the
dispersion law around these frequencies has the form
 \begin{align}\label{flatB}
&[\omega^c_{\vec k}/\omega^{}_0]|_{k_x\approx
0}\approx  0.3584 \nonumber\\
&+0.564\times 10^{-3}\cos(\sqrt{3}a k_y) +f(k_y) (a k_x)^2\ ,
 \end{align}
with $f(k_y)\sim 1$. These modes correspond to a horizontal
displacement of every second row in the configuration
of~\ref{fig:1}$c$. The small coefficient of the second term in
Eq.~(\ref{flatB}) implies that  different electron rows ``see"
each other as a continuous charge lines. This is surprising, since
the distance between the  rows is only $\sqrt{3}$-times larger
than the interval between electrons within each  row.

%\noindent
{\it 5. Squeezing the crystal.} The crystal structures of
Figs.~\ref{fig:1}$b$ and $1c$ have a preferential axis, violating
the $60^\circ$ symmetry of the triangular lattice. This opens the
possibility for a (slight) squeezing of the lattice, caused by an
interplay of the classical Coulomb repulsion and the quantum
vibrations ({\it cf.}  the strong squeezing in the case of a
short-range interaction~\cite{Berg12}).

The `density preserving' squeezing is defined as $\vR_i^{}
\rightarrow \tilde\vR_i^{} =[(1+\alpha)X^{}_{i},
Y^{}_{i}/(1+\alpha)]$, where $\alpha$ is a small but {\em finite}
parameter. We may also write the energy per electron in a crystal
as a power series in the Plank's constant,
$E(\alpha)=\epsilon_0(\alpha) +\hbar \, \epsilon_1(\alpha)+\cdots
$. The first term is the electron's electrostatic energy,
 \begin{align}\label{E01}
\epsilon_0(\alpha)&=({e^2}/{a})(c_0 +\alpha c_1 +\alpha^2 c_2
+\ldots)\ ,
\end{align}
and the second comes from the average zero-point energy,
 \begin{align}
%\label{E01}
%E_0(\alpha)&=({e^2}/{a})(c_0 +\alpha c_1 +\alpha^2 c_2 +\ldots)\ ,
%\nonumber\\
\epsilon_1(\alpha)&=\omega_0(d_0 -\alpha d_1 +\ldots)\ ,
 \end{align}
where $c_i$ and $ d_i$ are numerical coefficients. The
coefficients $c_0$ and $d_0$ [the latter is found in
Eq.~(\ref{threeOmega})] are of no interest. Since the triangular
lattice is the minimum of the electronic Coulomb
energy~\cite{Bonsall79}, one has $c_1\equiv 0$ and $c_2>0$. The
linear in $\alpha$ term in the quantum correction to the energy,
$d_1$, would also vanish due to the crystal symmetry in the
absence of the spin-orbit coupling, or e.g. for the
$120^\circ-$rotational symmetric configuration
Fig.~\ref{fig:1}$d$, but is allowed for the configurations
Figs.~\ref{fig:1}$b$ and $c$. Numerically we found $c_2\approx
0.527$, $d_1^b\approx 0.245$ and $\ d_1^c\approx 0.425$.
Minimizing the energy, $E(\alpha)$, results in
 \begin{align}
\alpha^{}_c= \fr{\hbar\omega^{}_{0}}{2 e^2/a}\fr{d^{}_1}{c^{}_2} =
\sqrt{\fr{\gamma a^{}_{\rm B}}{4 a}} \fr{d^{}_1}{c^{}_2}\sim
\sqrt{\fr{a_B}{a}}\ .
 \end{align}
The value of the squeezing parameter, $\alpha_{c}$, grows with the
density as $a^{-1/2} \sim n^{1/4}$. This increase is
limited, however,  by the inequality (\ref{regime}),

{\it 6. Summary.}  As we have shown, the possibility of the
electrons to occupy different minima in momentum space leads to a
complicated ground-state of the Wigner crystal. We found the
ground state by considering an effective Hamiltonian, which
accounts for a {\it single energetic} vibrational mode per
electron, leaving untouched a number of non-phonon excitations
from the low-energy sector of the full problem. These soft modes
would correspond to a small displacement of the electron wave
packet in momentum space along the circle of degenerate minima
(see Fig.~\ref{fig:1}$a$), or to  spin flips associated with the
$180^\circ$ jumps of the electron to the opposite side of the
circle of minima (see Refs.~\cite{Flambaum99} and \cite{Starykh08}
for a discussion of the spin structure of a Wigner crystal.)

Although we expect the true ground state of the crystal to be the
structure in Fig.~\ref{fig:1}$c$, other configurations may exist
as metastable states. If the configuration 1$b$ would be realized
experimentally, one will be able to probe the angular-dependent
plasmon modes, Eq.~(\ref{Omega}).

Finally, the existence of the spin-orbit dominated phase of the
Wigner crystal described in this paper requires the validity of
the inequality (\ref{regime}). This  may be rewritten as
$m^*\lambda^2\gg {\mathrm{Ry}}/r_s^{3/2}$, where
${\mathrm{Ry}}=m^* e^4/2\hbar^2\epsilon^2$ and $m^*$ and
$\epsilon$ are the effective mass and dielectric constant,
respectively. For InAs \cite{Grundler},   $m^*\lambda^2\approx
0.2{\rm meV}$ and ${\mathrm{Ry}}\approx 2.5 {\mathrm {eV}}$.
Assuming that in the presence of spin-orbit interactions the
crystal stability still requires the  large value of $r_s\sim 35$,
we expect our results to be always applicable for Wigner crystals
in such materials.

\begin{acknowledgments}

We  thank Y.~Imry, P. W.~Brouwer, E.~Bergholtz and M.~Schneider
for helpful discussions. The hospitality of the Institute for
Advanced Studies at the Hebrew University and the Albert Einstein
Minerva Center for Theoretical Physics  at the Weizmann Institute
are gratefully acknowledged. This work was supported by the
Alexander von Humboldt Foundation, the DFG grant RE 2978/1-1, the
Israeli Science Foundation (ISF), and the US-Israel Binational
Science Foundation (BSF).
\end{acknowledgments}

%\title{{\underline{Supplementary material for}}\\
%{\bf Wigner crystal of a two-dimensional electron gas \\
%with a strong spin-orbit interaction}}
%
%\author{P. G. Silverstrov}
%
%\author{O. Entin-Wohlman}
%
%\date{\today}
%
%\maketitle

\newpage

\begin{widetext}

\section{Appendix: Vibration spectrum for several electrons per unit cell}

We first show how to obtain the phonon spectrum belonging to
configuration $1c$ (Fig. 1 in the paper).  This is an example of a
lattice with two electrons per unit cell (which turns out to be
our best candidate for the ground state). We then sketch briefly
the derivation of the spectrum in the case of three electrons per
unit cell,  vibrating along three arbitrary directions.

\subsection{ The configuration of Fig. $1c$}

When there are two electrons within each unit cell, it is
convenient to introduce two sublattices, $a$ and $b$. The
effective vibration Hamiltonian for the configuration of Fig. $1c$
is then
\begin{align}
{\cal H}^{}_{\rm eff}&=\sum_{i\in a} \left( \frac{p_{x_i}^2}{2m}
+\frac{m\omega_{0}^2 x_i^2}{2}\right) +e^2\sum_{i<j\in a}
\frac{x_ix_j}{R_{ij}^3} \left( 1
-3\frac{X_{ij}^2}{R_{ij}^2} \right)\nonumber\\
&+\sum_{i\in b} \left( \frac{p_{y_i}^2}{2m} +\frac{m\omega_{0}^2
y_i^2}{2}\right) +e^2\sum_{i<j\in b} \frac{y_iy_j}{R_{ij}^3}
\left( 1 -3\frac{Y_{ij}^2}{R_{ij}^2} \right)-3e^2\sum_{i\in a,
j\in b} {x_iy_j}\frac{X_{ij} Y_{ij}}{R_{ij}^5}\ .
 \end{align}
Introduction of the creation/annihilation operators
 \begin{align}
a=\sqrt{\frac{m\omega^{}_{0}}{2\hbar}}(x+
\frac{i}{m\omega^{}_{0}}p_x), \ b
=\sqrt{\frac{m\omega^{}_{0}}{2\hbar}}(y+
\frac{i}{m\omega^{}_{0}}p_y)\ ,
 \end{align}
leads to the Hamiltonian
  \begin{align}
{\cal H}^{}_{\rm eff}&=\hbar\omega^{}_{0} \left( \sum_{i\in a}
(a_i^\dagger a^{}_i +\frac{1}{2}) +\sum_{i<j\in a}
v_{ij}^x (a^{}_i +a_i^\dagger)(a^{}_j +a_j^\dagger)\right.\nonumber\\
&\left. + \sum_{i\in b} (b_i^\dagger b^{}_i +\frac{1}{2})
+\sum_{i<j\in b} v_{ij}^y (b^{}_i +b_i^\dagger)(b^{}_j
+b_j^\dagger)\right. \left. +\sum_{i\in a,j\in b} w^{}_{ij}
(a^{}_i +a_i^\dagger)(b^{}_j +b_j^\dagger)\right)\ ,
  \end{align}
where
 \begin{align}
v_{ij}^x=\frac{a^3}{2\gamma R_{ij}^3}(1- 3X_{ij}^2/R_{ij}^2)\ , \
\ v_{ij}^y=\frac{a^3}{2\gamma R_{ij}^3}(1- 3Y_{ij}^2/R_{ij}^2)\ ,\
\ w^{}_{ij}=-3\frac{a^3 X_{ij} Y_{ij}}{2\gamma R_{ij}^5}\ .
 \end{align}
In Fourier space, this Hamiltonian takes the form  %
\begin{align}
{\cal H}^{}_{\rm eff}&= \hbar\omega^{}_{0} \sum_{\bf k} [(a_{\bf
k}^\dagger a_{\bf k}^{} +1/2) + \frac{v^x({\bf k})}{2} (a_{\bf
k}^{} +a_{-{\bf k}}^\dagger)(a_{-{\bf k}} ^{}+a_{\bf
k}^\dagger)\nonumber\\
&+ (b_{\bf k}^\dagger b_{\bf k}^{} +1/2) + \frac{v^y({\bf k})}{2}
(b_{\bf k}^{} +b_{-{\bf k}}^\dagger)(b_{-{\bf k}}^{} +b_{\bf
k}^\dagger) +w({\bf k}) (a_{\bf k}^{} +a_{-{\bf
k}}^\dagger)(b_{-{\bf k}}^{} +b_{\bf k}^\dagger)] \ ,
  \end{align}
with
 \begin{align}
 \label{Fourvx}
v^x({\bf k})=v^x(-{\bf k})=& \sum_{0\in a,i\neq 0\in a} v_{0i}^x
e^{i{\mathbf k} \cdot{\mathbf R_i}}\  ,\ \ v^y({\bf k})=v^y(-{\bf
k})= \sum_{0\in a,i\neq 0\in a} v_{0i}^y e^{i{\mathbf k}
\cdot{\mathbf
R_i}}\ ,\nonumber\\
& w({\bf k})=w(-{\bf k})= \sum_{0\in a,i\in b} w_{0i} e^{i{\mathbf
k}\cdot {\mathbf R_i}}\ .
 \end{align}
An important point to note here is that $v^y({\bf k}) \neq
v^x({\bf k})$. In the next step one decouples the two vibration
polarizations,
 \begin{align}
c_{\bf k}^{}=\cos(\tau^{}_{\bf k}) a{}_{\bf k} +\sin(\tau^{}_{\bf
k}) b_{\bf k}^{}\ ,\  \ d_{\bf k}^{} =-\sin(\tau ^{}_{\bf
k})a_{\bf k }^{}+\cos(\tau^{}_{\bf k}) b_{\bf k}^{}\ ,
 \end{align}
with $\tan (2\tau^{}_{\bf k})=2w({\bf k})/[v^x({\bf k})- v^y({\bf
k})]$, to obtain
 \begin{align}
{\cal H}^{}_{\rm eff}&=\hbar\omega^{}_{0} \sum_{\bf k} [(c_{\bf
k}^\dagger c_{\bf k}^{} +1/2) + \frac{v_1({\bf k})}{2} (c_{\bf
k}^{}
+c_{-{\bf k}}^\dagger)(c_{-{\bf k}}^{} +c_{\bf k}^\dagger)\nonumber\\
&
 + (d_{\bf k}^\dagger d_{\bf k}^{} +1/2) + \frac{ v_2({\bf k})}{2} (d_{\bf k}^{}
+d_{-{\bf k}}^\dagger)(d_{-{\bf k}}^{} +d_{\bf k}^\dagger)] \ ,
 \end{align}
where
 \begin{align}
v^{}_{1,2}({\bf k})= \frac{1}{2}[v^x({\bf k}) + v^y ({\bf
k})\pm\sqrt{(v^x({\bf k})-v^y({\bf k}))^2 +4w^2({\bf k})}]\ .
 \end{align}
Each polarization can be now Bogoliubov transformed exactly as is
carried out in the main text.

\begin{figure}
\includegraphics[width=12.cm]{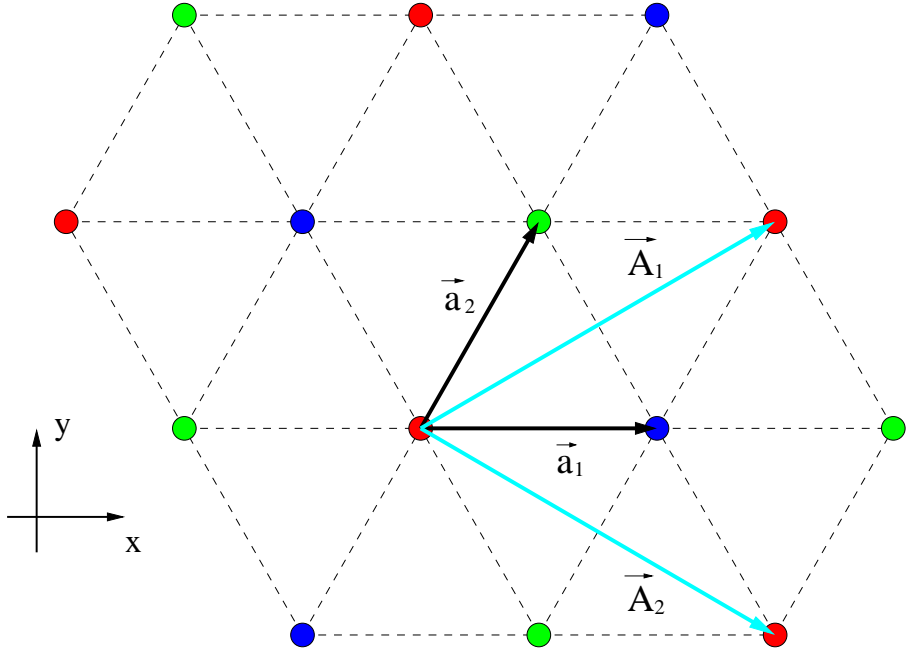}
\caption{ Definition of the basis vectors for the original
triangular lattice, ${\bf a}_1$ and ${\bf a}_2$, and for the
superlattice with three electrons per unit cell, ${\bf A}_1$ and
${\bf A}_2$. Electrons from the three different sublattices are
shown by different colors.
 } \label{fig3}
\end{figure}

\subsection{Phonon spectrum for superlattices with three electrons per unit
cell}

The triangular lattice of a usual Wigner crystal
%(see Fig. $1a$ in the main text)
is defined by two lattice vectors, ${\bf a}_1=(a,0)$ and ${\bf
a}_2=(a/2,\sqrt{3}a/2)$. The superlattice with three atoms per
unit cell
%(e.g.,   the configuration shown in Fig. $1d$)
is defined by the lattice vectors ${\bf A}_1=(3a/2,\sqrt{3}a/2)$
and ${\bf A}_2=(3a/2,-\sqrt{3}a/2)$, such that $|{\bf A}_1|=| {\bf
A}_2|=\sqrt{3}a$, as is shown in the figure. The electrons on this
lattice naturally form three sublattices whose sites  coordinates
are $i{\bf A}_1 +j{\bf A}_2$, $(a,0)+ i{\bf A}_1 +j{\bf A}_2$, and
$(-a,0)+ i{\bf A}_1 +j{\bf A}_2$. Here $i$ and $j$ are two
arbitrary integer numbers.

Let the displacement vectors for the three sublattices be ${\bf
u}_i$,  ${\bf v}_i$,  and ${\bf w}_i$, where ${\bf u}=u {\bf
n}_u$, ${\bf v}=v  {\bf n}_v$,  and ${\bf w}=w {\bf n}_w$, with
the three unit vectors $ {\bf n}_u$, ${\bf  n}_v$,  and ${\bf
n}_w$ pointing along the directions of the allowed vibrations for
each sublattice. For example, for the electronic configuration
shown in Fig. $1d$ of the main text one has $ {\bf n}_u= (0,1)$, $
{\bf n}_v= (\sqrt{3}/2,-1/2)$,  and ${\bf  n}_w=
(-\sqrt{3}/2,-1/2)$. However, our derivation below does not assume
any specific orientation of ${\bf n}_u $,  ${\bf n}_v$, and $ {\bf
n}_w $.

The Hamiltonian  takes the form
 \begin{align}
H=h^{}_u +h^{}_v +h^{}_w +V^{}_u +V^{}_v +V^{}_w +W^{}_{uv}
+W^{}_{uw} +W^{}_{vw}\ ,
 \end{align}
where,  for example,
 \begin{align}
h_u^{}=\sum_{i\in u} \left( \frac{p_{u_i}^2}{2m} +\frac{m\omega^2
u^2}{2} \right)\ ,
 \end{align}
and
 \begin{align}
V_u^{}&=e^2\sum_{i<j\in u}u^{}_i u^{}_j\left( \frac{1}{R_{ij}^3} -
3\frac{({\bf n}^{}_u\cdot {\bf R}_{ij})^2}{R_{ij}^5}\right) \ ,\nonumber\\
W^{}_{uv}&=e^2\sum_{i\in u,j\in v}u_i v_j\left( \frac{1}{R_{ij}^3}
- 3\frac{({\bf n}^{}_u\cdot {\bf R}^{}_{ij}) ({\bf n}^{}_v\cdot
{\bf R}^{}_{ij})}{R_{ij}^5}\right)\ .
 \end{align}

Let
 \begin{align}
u_i^{}=\sqrt{\frac{3}{N}}\sum_{\bf k}e^{i{\bf  k}\cdot {\bf
R}^{}_i }u^{}_{\bf k} \ , \ v^{}_i=\sqrt{\frac{3}{N}}\sum_{\bf
k}e^{i{\bf k}\cdot {\bf R}^{}_i} v^{}_{\bf k} \ , \
w^{}_i=\sqrt{\frac{3}{N}}\sum_{\bf k}e^{i{\bf k}\cdot {\bf  R}_i}
w^{}_{\bf k} \ ,
 \end{align}
where $N$ is the total number of electrons and ${\bf R}_j$ is the
true coordinate of the corresponding $N$-electron lattice site.
Similarly we introduce the Fourier transformed momenta $p_{u_{\bf
k}},p_{v_{\bf k}},p_{w_{\bf  k}}$, so that $[p_{\bf k}, x_{\bf
q}]=-i\delta_{{\bf k}+{\bf q}}$. It follows that
 \begin{align}
h_{u_{\bf k}}=\sum_{\bf k} \left( \frac{p_{u_{-{\bf k}}} p_{u_{\bf
k}}}{2m} +\frac{m\omega^2 u_{-{\bf k}}u_{\bf k}}{2} \right)\ ,
 \end{align}
 \begin{align}
V_u\rightarrow V_{u}^{({\bf k})}=\frac{e^2}{2}\sum_{\bf  k}
u^{}_{-{\bf  k}} u^{}_{\bf k}{\cal V}_{u}({\bf k}), \ \
% \ee
%\ {\rm where} \
% \bq
{\cal V}_{u}({\bf  k})= \sum_j e^{i{\bf  k}\cdot {\bf
R}^{}_{0j}}\left( \frac{1}{R_{0j}^3} - 3\frac{({\bf n}^{}_u\cdot
\bf{R}_{0j})^2}{R_{0j}^5}\right)\ ,
 \end{align}
 and
 \begin{align}
W_{uv}\rightarrow W_{uv}^{({\bf k})}&=e^2\sum_{\bf k} u^{}_{-{\bf
k}}
v^{}_{\bf k}{\cal W}_{uv}({\bf  k}) \ ,\nonumber\\
{\cal W}_{uv}({\bf  k})&= \sum_j e^{i{\bf  k}\cdot {\bf
R}^{}_{0j}}\left( \frac{({\bf n}^{}_u\cdot {\bf
n}^{}_{v})}{R_{0j}^3} - 3\frac{({\bf n}^{}_u\cdot {\bf
R}^{}_{0j})({\bf n}^{}_v\cdot {\bf R}^{}_{0j})}{R_{0j}^5}\right)\
.
 \end{align}

Next we perform a unitary rotation of the coordinates $u_{\bf k}$,
$v_{\bf  k}$,   and $w_{\bf k}$,  in order to diagonalize the
matrix
 \begin{align}
\frac{e^2}{2}\left(\begin{array}{cc} \ {\cal V}^{}_{u}(\bf k) \ \
{\cal
W}^{}_{uv}({\bf  k}) \ {\cal W}^{}_{uw}({\bf  k})\\
{\cal W}^{}_{vu}({\bf  k}) \ \ {\cal V}^{}_{v}({\bf  k}) \ \
{\cal
W}^{}_{vw}({\bf  k}) \\
{\cal W}^{}_{wu}({\bf  k}) \ {\cal W}^{}_{wv}({\bf  k}) \ \ {\cal
V}^{}_{w}({\bf  k})
\end{array}\right)\ ,
 \end{align}
which we do numerically. After that the Bogoliubov transformation
for each polarization is carried out similar to the way it is done
in the main text.

\end{widetext}

\end{document}